\documentclass[pra]{revtex4}
\usepackage{amssymb}
\usepackage{amsmath}
\usepackage{graphics}
\newcommand{\vc}[1]{\boldsymbol{#1}}

\begin{document}

\title{Hermitian spin-orbit  Hamiltonians  on a surface in orthogonal curvilinear coordinates: a  new practical approach}

\author{M. S. Shikakhwa}
\affiliation{Department of physics, University of Jordan,Amman 11942 Jordan
and Middle East Technical University Northern Cyprus Campus,Kalkanl\i, G\"{u}zelyurt, via Mersin 10, Turkey}
\author{N. Chair}
\affiliation{Department of physics, University of Jordan,Amman 11942 Jordan}


\begin{abstract}
The Hermitian Hamiltonian of a spin one-half particle with spin-orbit coupling (SOC) confined to a surface that is embedded in a three-dimensional space spanned by  a general Orthogonal Curvilinear Coordinate (OCC) is constructed. A gauge field  formalism, where the SOC is expressed as a non-Abelian $SU(2)$  gauge field is used. A new practical approach, based on the physical argument that upon confining the particle to the surface by a potential, then it is the physical Hermitian momentum operator transverse to the surface, rather than just the derivative with respect to the transverse coordinate  that should be dropped from the Hamiltonian.Doing so, it is shown that the Hermitian Hamiltonian for SOC is obtained with the geometric potential  and the geometric kinetic energy terms emerging naturally. The geometric potential is shown to represent a coupling between the transverse component of the gauge field and the mean curvature of the surface that replaces the coupling between the transverse momentum and the gauge field. The  most general Hermitian Hamiltonian with linear SOC on a general surface embedded in any 3D OCC system is reported. Explicit plug-and-play formulae for this Hamiltonian on the surfaces of a cylinder, a sphere and a torus are given. The formalism is applied to the  Rashba SOC in three dimensions (3D RSOC) and the explicit expressions for the surface Hamiltonians on these three geometries are worked out.
\end{abstract}

\maketitle

\section{Introduction}
There has been a revival of  interest in two-dimensional electron and hole systems with spin-orbit coupling (SOC) motivated by recent developments in spintronics. \cite{spintronics}. The magnetic field induced by the SOC may be tuned by varying the electric potential felt by the charge carriers in the crystal, thus providing a mechanism for spin current generation and manipulation. So far,two major types of SOC mechanisms have been receiving the most interset; the one resulting from the structure inversion asymmetry, or the Rashba SOC (RSOC)\cite{rashba}, while the second is due to the bulk inversion asymmetry and known as Dresselhaus SOC (DSOC)\cite{dresselhaus}. An approach to investigate SOC systems that has gained some popularity recently is the non-Abelian gauge field formulation of the SOC \cite{Jin.et.al-JPA,Dartora08,shikakhwa12}. The idea introduced in \cite{frohlich93} is based on noting that a linear SOC ;  $\beta\vc{p}\cdot(\vc{\sigma}\wedge\vc{E})$ can be  expressed in terms of  a non-Abelian gauge field $\vc{W}$ related to the electric field $\vc{E}$ generating the SOC as $-gW_i^a \sigma^a= 2m\beta\epsilon_{iaj}E_j$, ($\vc{p}$ is the momentum operator, $\sigma^i$'s are the Pauli spin matrices, and $\beta$ a constant related to the strength of the coupling), thus allowing one to write the Hamiltonian using the  covariant derivatives as in Eq.(\ref{a}) below. This approach has the advantage of allowing one to avail from the well-established full machinery of dealing with non-Abelian gauge fields; where ideas like gauge transformations, gauge invariance..etc. can be employed. Many authors have exploited this approach efficiently to reveal some interesting properties of 2D spin one-half systems with SOC \cite{Tokatly08,Tokatly and Sherman,chen&chang08,yang.et.al08,medina-epl08,berche graphene}. An interesting point in applying this approach to SOC systems is the fact that the gauge field $\vc{W}$ in this case  - unlike the case in particle physics- is related to a physical field as is evident from its definition. Therefore, a gauge transformation, for example, amounts to a change in a physical field configuration , so a gauge-symmetric Hamiltonian describes systems with different field configurations that are unitarily-equivalent and iso-spectral.\\
More recently, spin one-half particle systems with SOC in geometries other than planar started to attract attention as a result of the advances in nano technology which made it possible to synthesize objects like nano tubes and nano bubbles
 \cite{Romanov,Turshin,Son-Hsein 2010,Magaril99,Hernando,Jeong}. It has been known for sometime that attempting to write down the  SOC Hamiltonians for reduced geometries by blindly following the pragmatic approach, where one simply "freezes" one of the coordinates and writes the Hamiltonian accordingly by dropping the corresponding derivatives gives rise to non-Hermitian Hamiltonians  \cite{meijer,Lyanda-Geller}. Therefore, many authors \cite{Entin and Magaril,u shape,cheng,exact} turned to the so-called thin layer quantization approach \cite{Costa} to formulate the correct Hamiltonian on general curved geometries. This approach provides a systematic way of confining a system to a curved surface embedded in a full three dimensional  general curvilinear coordinate system by introducing a  confining potential to freeze the coordinate normal to the surface and decouple it from the surface Hamiltonian. The results, however, are in most cases applied to to conventional geometries like ring, sphere and cylinder which can be described using \textit{orthogonal }curvilinear coordinate systems.\\

The present work presents a new, simple and physics-based approach to address the problem of writing the Hamiltonian of a spin one-half particle with \textit{linear} SOC confined to a surface that is embedded in a general \textit{orthogonal } curvilinear coordinate(OCC) using the gauge field formulation of SOC. It is based on the simple observation that when the particle is squeezed to a surface by a confining potential, then one drops from the Hamiltonian the physical momentum transverse to he surface, rather than just the derivative of the transverse coordinate. In doing so,the surface Hamiltonian acquires the well-known geometric potential and is rendered Hermitian; the geometric kinetic energy term emerges naturally as well. Moreover, it is shown that  for a constant SOC the geometric potential is actually a coupling of the gauge field $\vc{W}$ to the mean curvature of the surface, thus providing a physical interpretation of this term.\\
In section 2, the general formalism is introduced, and the Hermitian Hamiltonian with any linear SOC on a surface embedded in a 3D space spanned by OCC system is derived. Three special geometries are then considered; a cylinder, a sphere and a torus. General expressions valid for any linear SOC for the Hermitian Hamiltonians on these surfaces are derived.Section 3 applies the formalism to the RSOC, where the Hermitian Hamiltonians on the surfaces of a cylinder and a sphere reported earlier in the literature are derived, and the Hamiltonian for the RSOC on the surface of a torus is worked out. Discussion and conclusions are given in section 4.\\
\begin{figure}[h!]
\includegraphics{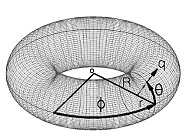}
\caption{Coordinates system $(\theta,\phi,q)$ near a torus surface. $R$ being the distance from the centre of the torus $O$ to the centre of the tube, $r$ is the radius of the tube and $q$ is the distance along the normal to the surface. }
\end{figure}

\section{General Hermitian Hamiltonian  on a surface in OCC's}

The Hamiltonian of a spin one-half particle of mass $m$ subject to a general linear spin-orbit coupling (SOC) has the general form:
\begin{align}\label{a}
    H=&\frac{\vc{p}^{2}}{2m}-\frac{g}{m}\vc{W}\cdot\vc{p} \\
  =&\frac{(\vc{p}-g\vc{W})^{2}}{2m}-\frac{g^{2}}{2m}\vc{W}\cdot\vc{W}
\end{align}
where $\vc{W}$ is a (three)two-component non-Abelian gauge field $\mathbf{W}=\vec{W}^a \sigma^{a}$ (or $\mathbf{W}_i=W_i^a \sigma^a$) with $i,a=1..3$. In this work
we will use $a,b,c,...$ to refer to group indices and $i,j,k...$ to refer to space indices. Evidently, the above Hamiltonian is not gauge-invariant due to
the presence of the gauge-symmetry breaking term $\frac{g^{2}}{2m}\vc{W}\cdot\vc{W}$ \cite{shikakhwa12}.
Recall that in a source-free region $\vc{\nabla}\cdot\vc{W}=0$ as a result of $\vc{\nabla}\times \vc{E}=0$. In a general orthogonal curvilinear coordinate (OCC) with coordinates $u_i$ , this Hamiltonian takes the explicit form:
\begin{equation}\label{OCC H}
H=\frac{-\hbar^2}{2m}\nabla^2+\frac{i\hbar g}{m}(\frac{1}{h_1}W_1\partial_1+\frac{1}{h_2}W_2\partial_2+\frac{1}{h_3}W_3\partial_3)
\end{equation}
$h_1 ,h_2$ and $h_3$ are the well-known \cite{Arfken} scale factors of the OCC's defined in terms of the derivatives of the position vector $\vc{r}$ as $\frac{\partial\vc{r}}{\partial u_i}=h_i\vc{\hat{e}}_i$, where $\vc{\hat{e}}_i$'s are the orthogonal unit vectors of the OCC system, and we use $\partial_i\equiv\frac{\partial}{\partial u_i}$  .
A surface embedded in the space spanned by the OCC is defined by setting  one coordinate that has  the dimensions of length; $u_3$ say,( thus has a  scale factor $h_3=1$ )  to a constant, e.g.$u_3=a$. Confining the particle to this surface can then be achieved by introducing a confining potential $V(u_3)$ - not shown in the above Hamiltonian-  that squeezes the particle into a thin layer around $a$ and then taking the limit $u_3arrow a$ . In a pragmatic approach to write down the Hamiltonian of a particle thus confined to the surface, the derivatives $\partial_3$ is set to zero in the above Hamiltonian  on the grounds that $u_3$ is a constant on the surface. The resulting Hamiltonian is :
\begin{equation}\label{surface H}
H^{surf.}=\frac{-\hbar^2}{2m}\nabla'^2+\frac{i\hbar g}{m}(\frac{1}{h_1}W_1\partial_1+\frac{1}{h_2}W_2\partial_2)
\end{equation}
where
\begin{equation}\label{laplacian prime}
 \frac{-\hbar^2}{2m}\nabla'^2=\frac{-\hbar^2}{2m}(\frac{1}{h_1 h_2 h_3})(\partial_1\frac{h_2 h_3}{h_1}\partial_1+\partial_2\frac{h_1 h_3}{h_2}\partial_2)
\end{equation}
is the Laplacian operator at the surface.
Evidently, the above Hamiltonian is  not Hermitian in the sense \cite{Griffiths} $\langle \Psi|H\Psi\rangle\neq\langle H\Psi|\Psi\rangle$ on the surface . Moreover, this approach fails to generate the geometric kinetic energy term that appears in the thin layer quantization \cite{Costa}.\\
 We now, introduce a new approach to the problem, that is both physically sound and practical, and gives a Hermitian Hamiltonian with the geometric kinetic energy term emerging naturally. We start by noting that the physical momentum in the OCC is not simply the non-Hermitian operator $-i\hbar\partial_3$ ! One can check easily that it is
\begin{equation}\label{hermitian p}
 p_3=-i\hbar(\frac{1}{h_3}\partial_3+\frac{1}{2h_1 h_2 h_3}\partial_3(h_1 h_2))
\end{equation}
that turns out to be Hermitian.  Now, noting the following two relations:
\begin{equation}\label{p3}
 \frac{i\hbar g}{m}(\frac{1}{h_3}W_3\partial_3)=\frac{- g}{m}W_3 p_3-\frac{i\hbar g}{m}W_3(\frac{1}{2h_1 h_2 h_3}\partial_3(h_1 h_2))
\end{equation},
\begin{equation}\label{p3 square}
 \frac{-\hbar^2}{2m}\nabla^2=\frac{p_3^2}{2m}+\frac{\hbar^2}{2m}(\frac{\partial_3^2(h_1 h_2)}{2h_1 h_2}-\frac{(\partial_3(h_1 h_2))^2}{(2h_1 h_2)^2})-\frac{\hbar^2}{2m}\nabla'^2
\end{equation}
we can cast the Hamiltonian, Eq.(\ref{OCC H}), in the form:
\begin{equation}\label{OCC H p3}
 H= \frac{p_3^2}{2m}+\frac{\hbar^2}{2m}(\frac{\partial_3^2(h_1 h_2)}{2h_1 h_2}-\frac{(\partial_3(h_1 h_2))^2}{(2h_1 h_2)^2})-\frac{\hbar^2}{2m}\nabla'^2-\frac{ g}{m}W_3 p_3-\frac{i\hbar g}{m}W_3(\frac{1}{2h_1 h_2 h_3}\partial_3(h_1 h_2))+\frac{i\hbar g}{m}(\frac{1}{h_1}W_1\partial_1+\frac{1}{h_2}W_2\partial_2)
\end{equation}
Now, we make the following argument: SOC has its physical origin as a coupling between the magnetic field induced in the rest frame of the moving particle by the electric field ( expressed in the original Cartesian Hamiltonian as a coupling with the gauge field;$\frac{g}{m}\vc{W}\cdot\vc{p}$ ). Therefore, as the confining potential squeezes the particle so that it is trapped at the surface, it is natural that it will not "feel" the magnetic filed due to its transverse motion since this motion is frozen. Setting $p_3$ to zero in the above Hamiltonian exactly achieves this by decoupling $W_3$ from the Hamiltonian. Moreover, the contribution of $p_3$ to the kinetic energy, i.e. the
$\frac{p_3^2}{2m}$ term also drops. The two surviving new terms in the above Hamiltonian; the second and the fifth, are respectively, the geometric kinetic energy \cite{Costa} and the geometric potential \cite{u shape,exact,Entin and Magaril}. Indeed, one can  check (see the appendix) that :
\begin{equation}\label{geometric K}
\frac{\hbar^2}{2m}(\frac{\partial_3^2(h_1 h_2)}{2h_1 h_2}-\frac{(\partial_3(h_1 h_2))^2}{(2h_1 h_2)^2})=-\frac{\hbar^2}{2m}(M^2-K),
\end{equation}
where $M$ and $K$ are the mean  and the Gaussian curvatures, respectively,
in complete agreement with the thin layer quantization \cite {Costa}. As for the geometrical potential,  it is easy to see that it is can be expressed as:
\begin{equation}\label{geometric potential}
  -\frac{i\hbar g}{m}W_3(\frac{1}{2h_1 h_2 h_3}\partial_3(h_1 h_2))=\frac{i\hbar g}{m}W_3 M
\end{equation}
which couples the gauge field to the mean curvature of the surface. Since the confining potential freezes the transverse component of the momentum and at the same time  introduces curvature into space by physically defining the surface, then we can view this term as "replacing" the coupling of the gauge field to the transverse momentum. This result supports the physical picture we are drawing. In the work \cite{Jensen}, which discusses confining a particle in an electromagnetic field to a surface,  similar term coupling the transverse component of the electromagnetic vector potential to the mean curvature was shown to appear. In that work, that term was undesirable as it represented a curvature contribution to the orbital magnetic moment of the particle. In our case, it does not carry such a meaning and its appearance is welcomed, even crucial as we now show. We first note that the condition $\vc{\nabla}\cdot\vc{W}=0$ implies for a \textit{constant Cartesian vector} $\vc{W}$ that:
\begin{equation}\label{div w}
W_3(\frac{1}{2h_1 h_2 h_3})\partial_3(h_1 h_2)=-W_3M=\frac{-1}{2h_1 h_2 h_3}(\partial_1(h_2 h_3 W_1)+\partial_2(h_2 h_3 W_2))=\frac{-1}{2}\vc{\nabla'}\cdot\vc{W'}
\end{equation}
where primes again indicates quantities on the surface and the absence of the third components. Using this, we can write the surface Hamiltonian as:

\begin{equation}\label{most general H}
H^{surf.}_{Herm.}=\frac{-\hbar^2}{2m}\nabla'^2-\frac{\hbar^2}{2m}(M^2-K)+\frac{i\hbar g}{m}(\frac{1}{h_1}W_1\partial_1+\frac{1}{h_2}W_2\partial_2)+\frac{i\hbar g}{2m}(\frac{1}{h_1h_2h_3})
(\partial_1(h_2h_3W_1)+\partial_2(h_1h_3W_2))
\end{equation}
The above equation provides the the most general Hermitian Hamiltonian with  a linear SOC on a surface in any OCC system. We stress here that the  last term in the above Hamiltonian plays a crucial role  in establishing Hermicity. Our equation (\ref{div w}) provides the physical origin of this term.
One can recast this Hamiltonian using covariant derivatives in the form
\begin{align}\label{H covariant}
 H^{surf.}_{Herm.}&=\frac{-\hbar^2}{2m}((\hat{u}_1(\frac{1}{h_1}\partial_1-\frac{ig}{\hbar}W_1)+(\hat{u}_2(\frac{1}{h_2}\partial_2-\frac{ig}{\hbar}W_2))^2-
 \frac{g^{2}}{2m}\vc{W'}\cdot\vc{W'}-\frac{\hbar^2}{2m}(M^2-K)\\\nonumber
   & =\frac{-\hbar^2}{2m}(\hat{u}_1D_1+\hat{u}_2D_2)^2-\frac{g^{2}}{2m}\vc{W'}\cdot\vc{W'}-\frac{\hbar^2}{2m}(M^2-K)\\\nonumber
  &=\frac{-\hbar^2}{2m}(\vc{D'}\cdot\vc{D'})-\frac{g^{2}}{2m}\vc{W'}\cdot\vc{W'}-\frac{\hbar^2}{2m}(M^2-K)
\end{align}
where we have defined the covariant derivatives $D_k=(\frac{1}{h_k}\partial_k-\frac{ig}{\hbar}W_k), k=1,2$, and the prime, again, denotes quantities on the surface and the absence of the 3-component.\\
 Table \ref{table1} provides the h-factors, $p_3$'s and the geometric kinetic energies (GKE)for cylindrical, spherical and the torus (see figure) coordinates. The results for the GKE's coincide with the well-known ones reported in the literature \cite{exact,ferrari}, and those for $p_r$ in cylindrical and spherical polar coordinates are the standard text book expressions for those operators. It is now straightforward to write down the Hermitian Hamiltonians in these three main OCC systems.
  \begin{table}[t]
  \centering
\begin{tabular}{|l|l|l|l|}
  \hline
  OCC&  h-factors  & $p_3$'s & GKE \\
   \hline
 Cylindrical& $u_1=\theta, u_2=z, u_3=r; h_\theta=r,h_z=h_r=1$ & $ p_r=-i\hbar(\partial_r+\frac{1}{2r})$ & $\frac{-\hbar^2}{8ma^2}$\\
   \hline
   Spherical& $u_1=\theta, u_2=\phi, u_3=r; h_\theta=r,h_\phi=r\sin\theta, h_r=1$ & $ p_r=-i\hbar(\partial_r+\frac{1}{r})$ & 0\\
     \hline
  Torus & $u_1=\theta, u_2=\phi, u_3=q; h_\theta=r+q,h_\phi=R+(r+q)\cos\theta, h_q=1$ & $ p_q=-i\hbar(\partial_q+\frac{(R+2(r+q)\cos\theta}{2(r+q)(R+r+q)\cos\theta})$ & $\frac{-\hbar^2 R^2}{8mr^2(R+r\cos\theta)^2}$\\
    \hline
\end{tabular}
  \caption{ The h-factors, $p_3$'s and GKE for cylindrical, spherical and torus coordinates. The expression for GKE is calculated by setting  $u_3=r=a$ at the surfaces of a cylinder and a sphere, and  $u_3=q=0$ at the surface of the torus}\label{table1}
\end{table}
 In cylindrical coordinates the Hermitian Hamiltonian on the  surface of a cylinder of radius $a$ is :
\begin{equation}\label{cyl}
H^{cyl.}_{Herm.}=\frac{-\hbar^2}{2m}(\hat{\theta}D_\theta+\hat{z}D_z)^2-\frac{g^{2}}{2m}(W_\theta^2+W_z^2)+\frac{-\hbar^2}{8ma^2}
\end{equation}
with,
\begin{equation}\label{cov cyl}
 D_\theta=(\frac{1}{a}\partial_\theta-\frac{ig}{\hbar}W_\theta),D_z=(\partial_z-\frac{ig}{\hbar}W_z).
\end{equation}
Dropping the $z-$dependence, one gets the Hermitian Hamiltonian on a ring of radius $a$.On the surface of a sphere of radius $a$,the Hamiltonian takes the form:
\begin{equation}\label{sph}
H^{sph.}_{Herm.}=\frac{-\hbar^2}{2m}(\hat{\theta}D_\theta+\hat{\phi}D_\phi)^2-\frac{g^{2}}{2m}(W_\theta^2+W_\phi^2)
\end{equation}
with,
\begin{equation}\label{cov sph}
 D_\theta=(\frac{1}{a}\partial_\theta-\frac{ig}{\hbar}W_\theta),D_\phi=(\frac{1}{a\sin\phi}\partial_\phi-\frac{ig}{\hbar}W_\phi).
\end{equation}
Finally, for a torus we have the Hermitian Hamiltonian on the surface of the torus ($qarrow 0$) as:
\begin{equation}\label{torus}
  H^{tor}_{Herm.}=\frac{-\hbar^2}{2m}(\hat{\theta}D_\theta+\hat{\phi}D_\phi)^2-\frac{g^{2}}{2m}(W_\theta^2+W_\phi^2)+\frac{-\hbar^2 R^2}{8mr^2(R+r\cos\theta)^2}
\end{equation}
with,
\begin{equation}\label{cov tor}
D_\theta=(\frac{1}{r}\partial_\theta-\frac{ig}{\hbar}W_\theta), D_\phi=(\frac{1}{R+r\cos\theta}\partial_\phi-\frac{ig}{\hbar}W_\phi)
\end{equation}
 Equations (\ref{cyl}-\ref{cov tor}) are the most general expressions for Hermitian Hamiltonians with any linear SOC on the surfaces of a cylinder, sphere and torus, and provide a sort of plug-and-play formulae to find the explicit Hamiltonians for any linear SOC on these surfaces. In the following section, we will be applying these to write down  the explicit Hamiltonians for the  case of 3D RSOC.

\section{Application to 3d rsoc}
We now apply the formalism to the isotropic 3D RSOC with  coupling strength $\alpha$ that was also considered in \cite{exact}.
The RSOC is casted in the form of a gauge field  as:
\begin{equation}\label{}
-\frac{g}{m}\vc{W}=\frac{\alpha}{\hbar}(\sigma_z-\sigma_y,\sigma_x-\sigma_z,\sigma_y-\sigma_x)
\end{equation}
Now, the question of applying the formalism developed in the previous section to this RSOC, is almost reduced to
expressing the above RSOC gauge field in cylilndrical,spherical, and torus coordinates. The procedure is straightforward and the results are summarized in Table 1.
Plugging the expressions for the various components of the gauge field into Eqs.(\ref{cyl}),(\ref{sph}) and (\ref{torus}) gives, respectively, the explicit exact Hermitian Hamiltonians of 3D RSOC on the surfaces of a cylinder,sphere and torus.\\

\begin{table}[t]
  \centering
\begin{tabular}{|l|l|}
  \hline
   OCC & $ W_1,W_2$ and $W_3 $ \\
   \hline
 Cylindrical & $ W_1=W_\theta=-(\frac{m\alpha}{g\hbar})(-\sigma_z(\cos\theta+\sin\theta)+\vc{\sigma}\cdot \hat{r})$ \\
  &$W_2= W_z=-(\frac{m\alpha}{g\hbar})(\sigma_y-\sigma_x)$\\
   &$W_3= W_r=-(\frac{m\alpha}{g\hbar})(\sigma_z(\cos\theta-\sin\theta)-\vc{\sigma}\cdot\hat\theta)$ \\
   \hline
   Spherical& $W_1=  W_\theta=-(\frac{m\alpha}{g\hbar})(\sigma_x(\cos\theta\sin\phi+\sin\theta)-\sigma_y(\cos\theta\cos\phi+\sin\theta)+\sigma_z(\cos\theta\cos\phi-\cos\theta\sin\phi)$\\
     &$W_2=W_\phi=-(\frac{m\alpha}{g\hbar})(\sigma_x\cos\phi+\sigma_y\sin\phi-\sigma_z(\sin\phi+\cos\phi))$ \\
     & $W_3=W_r=-(\frac{m\alpha}{g\hbar})(\sigma_x(\sin\theta\sin\phi-\cos\theta)+\sigma_y(\cos\theta-\sin\theta\cos\phi)+\sigma_z(\sin\theta\cos\phi-\sin\theta\sin\phi))$\\
     \hline
  Torus & $W_1=W_\theta=-(\frac{m\alpha}{g\hbar})(-\sigma_x(\sin\theta\sin\phi+\cos\theta)+\sigma_y(\sin\theta\cos\phi+\cos\theta)+\sigma_z(\sin\theta\sin\phi-\sin\theta\cos\phi))$\\
   & $ W_2=W_\phi= -(\frac{m\alpha}{g\hbar})(\sigma_x\cos\phi+\sigma_y\sin\phi-\sigma_z(\sin\phi+\cos\phi))$\\
    & $W_3=W_q= -(\frac{m\alpha}{g\hbar})(\sigma_x(\cos\theta\sin\phi-\sin\theta)+\sigma_y(\sin\theta-\cos\theta\cos\phi)+\sigma_z(\cos\theta\cos\phi-\cos\theta\sin\phi))$\\
    \hline
\end{tabular}
  \caption{ The components of the RSOC gauge field $W_1,W_2 $ and $W_3$ for the major OCC's}\label{table}
\end{table}
 So,we get the explicit Hermitian Hamiltonian on the surface of a cylinder,

\begin{align}\label{exp cyl}
 H^{cyl.}_{R}=&\frac{-\hbar^2}{2m}(\hat{\theta}(\frac{1}{\theta}\partial_\theta+\frac{im\alpha}{\hbar^2}(-\sigma_z(\sin\theta+\cos\theta)+\vc{\sigma}\cdot\hat{r}))+\hat{z}(\partial_z+\frac{im\alpha}{\hbar^2}(\sigma_y-\sigma_x)))^2
-\frac{g^2}{2m}(W_\theta^2+W_z^2)+\frac{-\hbar^2}{8ma^2} \\\nonumber
 =&\frac{-\hbar^2}{2m}(\frac{1}{a^2}\partial_\theta^2+\partial_z^2)+\frac{i\alpha}{a}(\sigma_z(\sin\theta+\cos\theta)- \vc{\sigma}\cdot\hat{r})\partial_\theta
 +i\alpha(\sigma_x-\sigma_y)\partial_z+\frac{i\alpha}{2a}(\sigma_z(\cos\theta-\sin\theta)-\vc{\sigma}\cdot\hat{\theta})+\frac{-\hbar^2}{8ma^2},
\end{align}
 on the surface of a sphere,
\begin{eqnarray}\label{exp sph}
H^{sph.}_{R}&=&\frac{-\hbar^2}{2m}(\hat{\theta}\frac{1}{a}\partial_\theta+\frac{im\alpha}{\hbar^2}(\sigma_x(\cos\theta\sin\phi+\sin\theta)-\sigma_y(\cos\theta\cos\phi+\sin\theta)+
\sigma_z(\cos\theta\cos\phi-\cos\theta\sin\phi)). \\\nonumber
&+&\hat{\phi}(\frac{1}{a\sin\theta}\partial_\phi+\frac{im\alpha}{\hbar^2}(\sigma_x\cos\phi+\sigma_y\sin\phi-\sigma_z(\sin\phi+\cos\phi)))^2-\frac{g^{2}}{2m}(W_\theta^2+W_\phi^2) \\\nonumber
 &=& \frac{-\hbar^2}{2m}(\frac{1}{a^2\sin\theta}\partial_\theta(\sin\theta\partial_\theta)+\frac{1}{a^2\sin^2\theta}\partial_\phi^2)-\frac{i\alpha}{a}(\sigma_x(\cos\theta\sin\phi+\sin\theta)-\sigma_y(\cos\theta\cos\phi+\sin\theta)\\\nonumber
&-&\sigma_z(\cos\theta\cos\phi-\cos\theta\sin\phi))\partial_\theta-\frac{i\alpha}{a\sin\theta}(\sigma_x\cos\phi+\sigma_y\sin\phi-\sigma_z(\sin\phi+\cos\phi))\partial_\phi\\\nonumber
&+&\frac{i\alpha}{2}(\sigma_x(\sin\theta\sin\phi-\cos\theta)+\sigma_y(\cos\theta-\sin\theta\cos\phi)+\sigma_z(\sin\theta\cos\phi-\sin\theta\sin\phi))(\frac{2}{a})
\end{eqnarray}
and finally on the surface of a torus:
\begin{eqnarray}\label{exp tor}
   H^{tor.}_{R}&
   =& \frac{-\hbar^2}{2m}(\hat{\theta}(\frac{1}{r}\partial_\theta+\frac{im\alpha}{\hbar^2}(-\sigma_x(\sin\theta\sin\phi+\cos\theta)+\sigma_y(\sin\theta\cos\phi+\cos\theta)+\sigma_z(\sin\theta\sin\phi-\sin\theta\cos\phi)).\\\nonumber
 &+&\hat{\phi}(\frac{1}{R+r\cos\theta}\partial_\phi+\frac{im\alpha}{\hbar^2}(\sigma_x\cos\phi+\sigma_y\sin\phi-\sigma_z(\sin\phi+\cos\phi))^2-\frac{g^{2}}{2m}(W_\theta^2+W_\phi^2)\\\nonumber
 &+&\frac{-\hbar^2 R^2}{8mr^2(R+r\cos\theta)^2}\\\nonumber&=& \frac{-\hbar^2}{2m}\big(\frac{1}{r^{2}}\partial_\theta^{2} -\frac{\sin\theta}{r(R+r\cos\theta)}\partial_\theta+\frac{1}{(R+r\cos\theta)^{2}}\partial_\phi^{2}\big)+i\alpha\big(\sigma_x(\sin\theta\sin\phi+\cos\theta)-\sigma_y(\sin\theta\cos\phi+\cos\theta)\\\nonumber&-&\sigma_z(\sin\theta\sin\phi-\sin\theta\cos\phi)\big)\frac{1}{r}\partial_{\theta}-\frac{i\alpha}{(R+r\cos\theta)}\big(\sigma_x\cos\phi+\sigma_y\sin\phi-\sigma_z(\sin\phi+\cos\phi)\big)\partial_{\phi}\\\nonumber
 &-&\frac{i\alpha(R+2r\cos\theta)}{2r(R+r\cos\theta)}\big(\sigma_x(\sin\theta-\cos\theta\sin\phi)+\sigma_y(\cos\theta\cos\phi-\sin\theta)+\sigma_z(\cos\theta\sin\phi-\cos\theta\cos\phi)\big)\\\nonumber
 &+&\frac{-\hbar^2 R^2}{8mr^2(R+r\cos\theta)^2}
\end{eqnarray}
The Hamiltonian for a ring of radius $a$ can be obtained from Eq.(\ref{exp cyl}) by dropping the derivatives with respect to $z$. Eqs.(\ref{exp cyl}) and (\ref{exp sph}) - as expressed in their expanded forms after the second equality signs- are exactly, Eq.(16) and (19), respectively, in reference \cite{exact}. Eq.(\ref{exp tor}) is reported for the first time to the best of our knowledge. The terms linear in $\alpha$ and not coupled to the derivatives in each of the above equations are the  geometric potentials reported in this and other references\cite{Entin and Magaril,u shape,cheng}.

\section{conclusions}
We have proposed a new physical practical approach to construct the Hermitian Hamiltonian of a spin one-half particle with linear SOC confined to a surface embedded in 3D in any OCC system within the gauge field formulation of SOC. The approach is based on the simple argument that as the particle is confined o a surface by a potential, then it is the transverse physical Hermitian momentum operator that needs to be dropped from the Hamiltonian rather than just the derivative with respect to the transverse coordinate as used to be done in the conventional practical approach to the problem. The approach, not only generates the geometric potential that renders the Hamiltonian Hermitian as well as the geometric kinetic energy term, it also provides a new physical interpretation of this geometric potential as a coupling of the transverse component of the gauge field to the mean curvature of the surface. As such, this approach substitutes the old practical approach to confine a particle to a surface embedded in 3D space spanned by an OCC system based on simply dropping the derivative with respect to the transverse coordinate from the Hamiltonian.It also can be considered as  providing a physical interpretation for the very general and rigorous thin-layer quantization procedure.\\
 The development is simple, transparent and elegant: The main strength is that given any linear SOC Hamiltonian, we know now how to write down its corresponding Hermitian version  , it is just Eq.(\ref{most general H}) . This last Hermitian Hamiltonian is the most general Hermitian Hamiltonian for any linear SOC on a surface in any OCC system. The Hermitian Hamiltonians for any linear SOC on the three surfaces of practical value; a cylinder, a sphere and a torus were provided, which furnish a plug-and-play formulae for any linear SOC. The formalism is applied to the experimentally important 3D RSOC, where the Hermitian Hamiltonians on the surfaces of a sphere and a cylinder reported earlier by other methods are reproduced, and the Hermitian Hamiltonian on the surface of a torus is worked out. Closing, we note that the present approach can be easily applied to other interactions, e.g. a particle coupled to a $U(1)$ vector potential (electromagnetic field). The applicability of the approach to  general curvilinear coordinates - not necessarily orthogonal- should not be difficult either, but still needs to be carefully analyzed.  These questions are under current investigation.
\section{appendix }
 In this appendix we give derivations to Eq.(\ref{geometric K}) and
Eq. (\ref{geometric potential}) using classical differential Geometry \cite{Forsyth, Willmore}.  To carry out our derivations, it is  convenient  to use notations from  differential geometry of curved surfaces $\vc{r}=\vc{r}(u_{1}, u_{2}) $ namely, $h_{1}=\sqrt{E}$,  $h_{2}=\sqrt{G}$  and set $h_{3}=1$ and hence the first fundamental form on the surface is $$ds^{2}=h_{1}du_{1}^{2}+h_{2}du_{2}^{2}= Edu_{1}^{2}+Gdu_{2}^{2},$$ while the second fundamental form is $$Ldu_{1}^{2}+Ndu_{2}^{2}. $$
Here, $ L=-\frac{\partial{\vc {N}}}{\partial u_{1}}.\frac{\partial\vc{r}}{\partial u_{1}},$  and $N=-\frac{\partial{\vc {N}}}{\partial u_{2}}.\frac{\partial\vc{r}}{\partial u_{2}},$ $\vc {N} $ being the normal vector to the surface. The left-hand side of Eq. (\ref{geometric K}), can be written as
 \begin{equation}\label{geometric KA}
\frac{\hbar^2}{2m}\Big(\frac{1}{4h_{1}^2 h_{2}^2}\Big(2h_{1}h_{2}^{2}(\partial_3^2h_1)+2h_{1}^{2}h_{2}(\partial_3^2h_2)+2h_{1}h_{2}(\partial_3h_1)(\partial_3h_2)-h_{2}^{2}(\partial_3h_1)^{2}-h_{1}^{2}(\partial_3h_2)^{2}\Big)\Big).
\end{equation}
The above expression can be simplified using the Lame's Equations \cite{Forsyth} that are reduced to  $\partial_3^2h_1=0 $ and $\partial_3^2h_2=0 $. Furthermore, using the fact that $ L=-\frac{h_{1}}{h_{3}}(\partial_3h_1)$ and $ N=-\frac{h_{2}}{h_{3}}(\partial_3h_2)$, then Eq. (\ref{geometric KA}) reads
\begin{eqnarray}\label{geometric KAA}
\frac{\hbar^2}{2m}\Big(\frac{1}{4h_{1}^2 h_{2}^2}\Big(2h_{1}h_{2}(\partial_3h_1)(\partial_3h_2)-h_{2}^{2}(\partial_3h_1)^{2}-h_{1}^{2}(\partial_3h_2)^{2}\Big)\Big)&=&-\frac{\hbar^2}{2m}\Big(\frac{L^{2}}{4E^{2}}+\frac{N^{2}}{4G^{2}}-\frac{2LN}{4EG}\Big)\nonumber\\&=&-\frac{\hbar^2}{2m}\Big( M^{2}-K\Big).
\end{eqnarray}
Here, we used $ M=\frac{EN+GL}{2EG}$, $ K=\frac{LN}{2EG}$ corresponding to the mean curvature and curvature of the surface in the orthogonal curvilinear coordinates. We now move to derive Eq. (\ref{geometric potential}). The left-hand side of the latter can be written as follows
\begin{eqnarray}\label{geometric potentialA}
  -\frac{i\hbar g}{2m}W_3\Big(\frac{h_{1}(\partial_3h_1)}{h_1^{2} h_3}+\frac{h_{2}(\partial_3h_1)}{h_2^{2} h_3}\Big)&=&\frac{i\hbar g}{m}W_3\Big(\frac{LG+NE}{2EG}\Big)\nonumber\\&=&\frac{i\hbar g}{m}W_3 M
\end{eqnarray}


\begin{thebibliography}{0}
\expandafter\ifx\csname natexlab\endcsname\relax\def\natexlab#1{#1}\fi
\expandafter\ifx\csname bibnamefont\endcsname\relax
  \def\bibnamefont#1{#1}\fi
\expandafter\ifx\csname bibfnamefont\endcsname\relax
  \def\bibfnamefont#1{#1}\fi
\expandafter\ifx\csname citenamefont\endcsname\relax
  \def\citenamefont#1{#1}\fi
\expandafter\ifx\csname url\endcsname\relax
  \def\url#1{\texttt{#1}}\fi
\expandafter\ifx\csname urlprefix\endcsname\relax\def\urlprefix{URL }\fi
\providecommand{\bibinfo}[2]{#2}
\providecommand{\eprint}[2][]{\url{#2}}

\end{thebibliography}


\begin{thebibliography}{99}
\bibitem{spintronics} I. \v{Z}uti\'{c}, J. Fabian and S. Das Sarma, Rev. Mod. Phys.
    \textbf{76}, 323 (2004).


\bibitem{rashba} E. I. Rashba, Sov. Phys. Solid State \textbf{2}, 1109 (1960).

\bibitem{dresselhaus} G. Dresselhaus, Phys. Rev. \textbf{100}, 580 (1955).

\bibitem{Jin.et.al-JPA} P-Q. Jin, Y-Q. Li and F-C. Zhang, J. Phys. A: Math. Gen. \textbf{39}, 7115 (2006).

\bibitem{Dartora08} C. A. Dartora and G. G. Cabrera, Phys. Rev. B \textbf{78}, 012403 (2008).

\bibitem{shikakhwa12} M.S.Shikakhwa,S. Turgut, N.K.Pak,  J. Phys. A: Math. Gen. \textbf{45}, 105305 (2012).

\bibitem{frohlich93} J. Fr\"{o}hlich and U. M. Studer, Rev. Mod. Phys. \textbf{65}, 733 (1993).


\bibitem{Tokatly08} I. V. Tokatly, Phys. Rev. Lett. \textbf{101}, 106601 (2008).

\bibitem{Tokatly and Sherman} I. V. Tokatly, E.Ya.Sherman, Ann.Phys.\textbf{325}, 1104 (2010).


\bibitem{chen&chang08} S-H. Chen and C-R. Chang, Phys. Rev. B \textbf{77}, 045324 (2008).

\bibitem{yang.et.al08} J-S. Yang, X-G. He, S-H. Chen and C-R. Chang, Phys. Rev. B \textbf{78}, 085312 (2008).

\bibitem{medina-epl08} E. Medina, A. Lopez and B. Berche, Eur. Phys. Lett. \textbf{83}, 47005 (2008).

\bibitem{berche graphene} Berche, B., Bolívar, N., López, A., Medina, E. , The European Physical journal B \textbf{88}, 198 (2015).

\bibitem{Romanov} Magarill, L.I. and Romanov, D.A. and Chaplik, A.V,JETP Lett. \textbf{64 }, 460 (1996).

\bibitem{Turshin} M. Trushin and J. Schliemann,New Journal of Physics \textbf{9 }, 346 (2007).

\bibitem{Son-Hsein 2010} C-L.Chen et. al. ,J. App. Phys. \textbf{108 }, 033715 (2010).

\bibitem{Magaril99} L.I.Magarill, A.V Chaplik,JETP \textbf{88 }, 815 (1999).

\bibitem{Hernando} D.Huertas-Hernando, F.Guinea, A.Brataas, Phys. Rev. B\textbf{74 }, 155426 (2006).

\bibitem{Jeong} J-S. Jeong, H-W. Lee   Lee, Phys. Rev. B\textbf{80 }, 075409 (2009).

\bibitem{meijer} F. E. Meijer, A. F. Morpurgo, T. M. Klapwijk, Phys. Rev. B\textbf{66 }, 033107 (2002).

\bibitem{Lyanda-Geller} A. G. Aronov, Y. B. Lyanda-Geller,  Phys. Rev. B\textbf{70 }, 343 (1993).

\bibitem{Entin and Magaril} M. V. Entin, L. I. Magarill, Phys. Rev. B\textbf{64 }, 085330 (2001).

\bibitem{u shape} M-H. Liu et. al., Phys. Rev. B \textbf{84}, 085307 (2011).

\bibitem{cheng} T-C. Cheng, J-Y. Chen, and C-R. Chang, Phys. Rev. B \textbf{84}, 214423 (2011).

\bibitem{exact} J-Y. Chang, J-S. Wu, and C-R. Chang Phys. Rev. B \textbf{87 }, 174413 (2013).

\bibitem{Costa} R. C. T. da Costa, Phys. Rev. A \textbf{23}, 1982 (1981).


\bibitem{Arfken} G. Arfken, J.Weber \textit{Mathematical Methods for Physicists, sixth edition},
    Elsevier, 2005.

\bibitem{Griffiths} D. Griffiths \textit{Introduction to Quantum Mechanics, second edition},
    Pearson Education limited, 2014.
\bibitem{ferrari} G. Ferrari and G.Coughi, Phys. Rev. Lett. \textbf{100}, 240403 (2008).

\bibitem{Jensen} B.Jensen and R.Dandoloff Phys. Rev. A \textbf{80}, 052109 (2009).

\bibitem{Forsyth} A. R. Forsyth, Lectures on the differential geometry of curves and surfaces, second eddition, Cambridge University Press, 1920.
\bibitem{Willmore} T. J. Willmore, An Introduction to Differential Geometry, Oxford University Press 1959.
















\end{thebibliography}
\end{document}